\documentclass[twocolumn,amssymb, nobibnotes, showpacs, superscriptaddress, aps, prd]{revtex4}
\usepackage{amsmath}
\usepackage{epsfig}

\begin{document}
\title{Impact of photo-assisted collisions on superradiant light scattering with Bose condensates}
\author{Xinyu Luo}
\affiliation{Institute of Physics, Chinese Academy of Sciences, Beijing 100190, China}
\author{Kuiyi Gao}
\affiliation{Institute of Physics, Chinese Academy of Sciences, Beijing 100190, China}
\author{L. Deng}
\affiliation{Physical Measurement Laboratory, National Institute of Standards \& Technology, Gaithersburg, Maryland USA 20899}
\author{E.W. Hagley}
\affiliation{Physical Measurement Laboratory, National Institute of Standards \& Technology, Gaithersburg, Maryland USA 20899}
\author{Ruquan Wang}
\affiliation{Institute of Physics, Chinese Academy of Sciences, Beijing 100190, China}
\date{\today}

\begin{abstract}

We present experimental evidence supporting the postulation that the secondary effects of light-assisted collisions are the main reason that the superradiant light scattering efficiency in condensates is asymmetric with respect to the sign of the pump-laser detuning. Contrary to the recent experimental study, however, we observe severe and comparable heating with all three pump-laser polarizations. We also perform two-color, double-pulse measurements to directly study the degradation of condensate coherence and the resulting impact on the superradiant scattering efficiency.

\end{abstract}

\maketitle
The astonishingly efficient suppression \cite{exp} of superradiance 
\cite{inouye,ketterle,moore,li,piovella,pu,bonifacio,benedek,zobay,yu} observed with blue-detuned, linearly polarized pump fields has triggered further experimental and theoretical
investigations.  Although various suppression mechanisms based on single-atom light scattering have been studied \cite{lu1}, to date none are capable of explaining the magnitude of the suppression. Recently, a light-assisted collision model \cite{muller} was proposed that explains the suppression with blue detunings as a consequence of radiation trapping of photons emitted during the collisional process. These scattered photons are predicted to be red shifted near the single-photon resonance and therefore substantially heat the gas.  While we believe this model is correct, their work raised additional questions because contrary to their own calculation, Ref. \cite{muller,muller2} reports no evidence of heating and only a very weak impact (no suppression) on the superradiant process.   
These experimental findings are in stark contrast to the astonishingly efficient suppression of superradiant scattering with blue-detuned pumps reported previously \cite{exp}. In this paper we present additional experimental studies to clarify these differences.

\begin{figure}
\centering
\includegraphics[angle=90,width=3.2in]{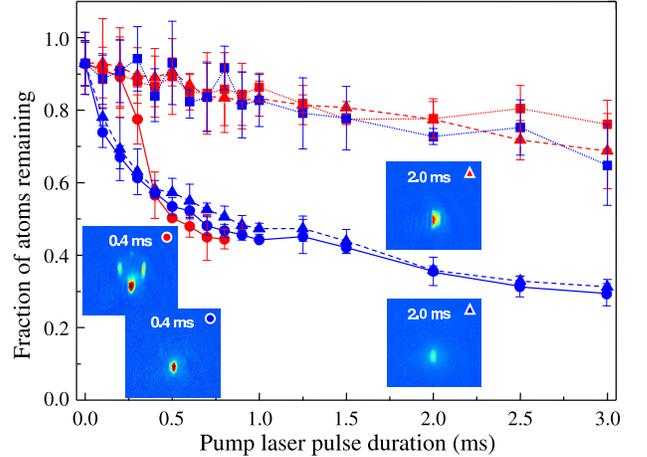} 
\caption{Fraction of atoms remaining in the condensate ($0\,\hbar k$) as a function of $\tau$ with $R$ = 95 Hz, $|\delta|$/$2\pi$ = 10 GHz, and TOF = 15 ms. Red (blue) color corresponds to a negative (positive) detunings. Squares: parallel polarization applied 5 ms after turning magnetic trap off.  Triangles: parallel polarization applied with magnetic trap on.  Solid circles: perpendicular polarization applied with magnetic trap on. Throughout this manuscript, typical error bars are one standard deviation unless otherwise indicated.}   
\end{figure}
\vskip 5pt
We begin by investigating atom loss from a $^{87}$Rb condensate using a method analogous to that employed in previous photo-association studies with magneto-optical traps \cite{lett}. Here the pump laser is linearly polarized and propagates along the short axis of the condensate, as in the original pioneering study Ref. \cite{ketterle}, and the D2 line ($5S_{1/2}\rightarrow 5P_{3/2}$) is excited. After applying a single, far-detuned laser pulse of selected detuning ($\delta$), duration ($\tau$), and Rayleigh scattering rate ($R$) to the condensate, we turn off the magnetic trap and employ Time-of-Flight (TOF) absorption imaging to analyze the atoms remaining within our field of view. Figure 1 shows atom loss out of the zero-momentum condensate as a function of $\tau$ for different experimental conditions. Note that both parallel and perpendicular polarizations are expected to generate the usual spontaneous Rayleigh scattering. In addition, a perpendicular polarization is also expected to result in the buildup of significant coherent Rayleigh scattering which drives the superradiant growth process. We found that if the trapping potential is turned off to allow the atomic density to drop to a sufficiently low level before applying the pulse, the observed rate of atom loss with a parallel polarization for both red- (red squares) and blue-detuned (blue squares) pumps is indeed consistent with that expected only from spontaneous Rayleigh scattering. However, when a blue-detuned pump is applied with the trap on, both parallel (blue triangles) and perpendicular polarizations (blue circles) result in a rate of atom loss that is much faster than that expected from spontaneous Rayleigh scattering, even though no superradiantly scattered atoms are observed with either polarization.  In contrast, if a perpendicularly-polarized, red-detuned pump is applied when the trap is on, matter-wave superradiant scattering is clearly visible and these data (red circles) display a threshold behavior heralding the high gain superradiant growth region that results in significant depletion of the condensate. The in-trap data with blue-detuned pump light (blue triangles and circles) indicate that for long pump times the light-assisted collision rate drops considerably during the duration of the pulse.  This occurs because the density of the gas drops in time due to heating, and the lower density causes a corresponding drop in the photo-induced collision rate and radiation trapping. Therefore the loss rate, as evidenced by the asymptotic slopes of the curves in Fig. 1, returns to the usual single-particle Rayleigh scattering rate.

\begin{figure}
\centering
\includegraphics[angle=90,width=3.2in]{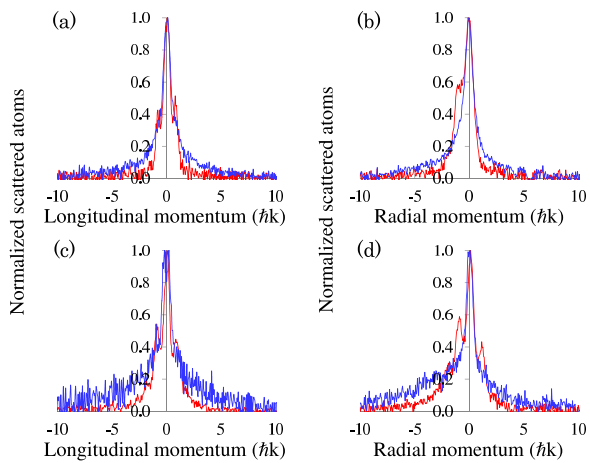} 
\caption{Normalized longitudinal (a) and transverse (b) momentum distributions of the atoms under red- and blue-detuned excitation for $\tau$ = 500 $\mu$s (a,b) and $\tau$ = 30 $\mu$s (c, d). Here, TOF = 15 ms, $R\approx$ 2 kHz, and $\delta/2\pi=\mp 2$ GHz for red/blue detunings, respectively.  Notice that a large fraction of the atoms acquire a momentum $>5\hbar k$, a feature observed for all pulse lengths studied.}   
\end{figure}
\vskip 5pt
To further understand the abnormally fast loss rate we analyzed the momentum spread of the atoms remaining within our 100 $\mu$m depth of field after applying blue- and red-detuned laser pulses. The spectra in Fig. 2 show that a substantial number of atoms acquired a momentum that was much larger than that expected from a first-order, two-photon Raman process. 
We note that because atoms participating in a photo-assisted collision acquire a very large velocity when they disassociate, they are not visible after the TOF and therefore do not directly contribute to the observed momentum distribution.   
In general, for the detunings we studied the terminal velocity of these atoms will be on the order of several m/s, and they will therefore leave our field of view in $\sim$ 100 $\mu$s. However, this direct loss of atoms from light-assisted collisions is small ($<1\%$). 

\vskip 5pt

It is well known \cite{suominen,walter} that during photo-assisted collision events with blue-detuned light that a large kinetic energy transfer to the constituent atom pair occurs and that the scattered/emitted photons are necessarily red-shifted closer to the single-photon resonance. As pointed out in Ref. \cite{muller}, if the detuning is very close to resonance then the photons will be radiation trapped and substantially heat the condensate. The radiation-trapped photons will then necessarily impact the superradiant scattering efficiency to a much greater degree than the direct loss of atoms from light-assisted collisions. Our data support the prediction of significant radiation trapping and heating, but are in stark contrast with the data presented in Ref. \cite{muller} where negligible heating and only a very weak impact on the superradiant threshold were reported.  Even the authors of Ref. \cite{muller} stated that their experimental observations did not agree with their calculations which were based on the theoretical framework of light-assisted collisions \cite{burnett}.
  
\begin{figure}
\centering
\includegraphics[angle=90,width=3.25in]{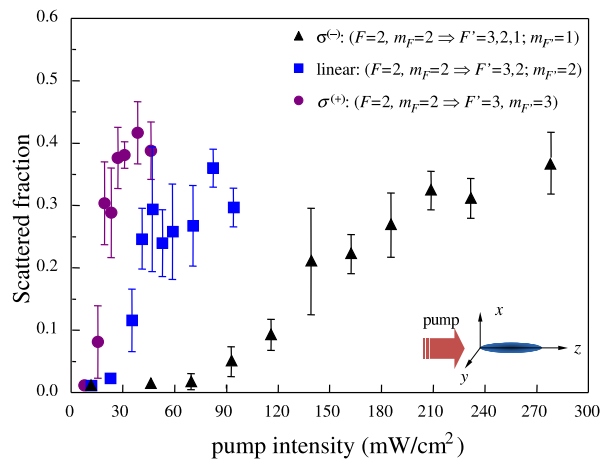} 
\caption{Plot of scattered fraction as a function of pump laser intensity for three laser polarizations with $\delta/2\pi=-5$ GHz.  The pump travels along the long axis as in Ref.\cite{muller}  The data agree with the corresponding estimates for a $2-$ ($\sigma^{(+)}$), a $3-$ (linear), and a $4-$ ($\sigma^{(-)}$) level systems.  In all cases, severe heating was observed.}  
\end{figure}
\vskip 5pt
It should be pointed out that Ref. \cite{muller} excited the D1 line of $^{87}$Rb using a circularly polarized $\sigma^{(-)}$ pump field that propagated along the magnetic field quantization axis (long axis of the condensate). However, the data reported in Ref. \cite{exp} and the data shown in Figs. 1 \& 2 were obtained using the D2 line of $^{87}$Rb with a linearly polarized pump field that propagated along the short axis of the condensate.  The D1 line experiment results in a true two-level system whereas the D2 line excitation of $^{87}$Rb leads to more complex multi-level excitations because the pump field propagates perpendicular to the magnetic quantization axis.  
\vskip 5pt
To further clarify this we carried out a series experiments that systematically investigated the heating effect using linear and $\sigma^{(\pm)}$ polarized pump light propagating along the long axis of the condensate. For D2 line excitation from a ground state of $F=2$, $m_F=+2$, pumping with $\sigma^{(+)}$, linear, and $\sigma^{(-)}$ results in a true $2-$, $3-$, and $4-$ level system, respectively.  We found that contrary to the D1 line data \cite{muller}, all measurements using a blue-detuned pump with different polarizations resulted in severe heating that is entirely consistent with the predictions of the light-assisted collision theory.  Figure 3 is a plot of the scattered fraction as a function of laser intensity for all three pump polarizations. These data agree with what is expected for the corresponding $2-$, $3-$, and $4-$ level systems, and the suppression of scattering and heating with blue detuned light was comparable for all three laser polarizations. We note that the resonant dipole-dipole coupling coefficient $C_3$ for $S+P_{3/2}$ (D2 line) is different from that of $S+P_{1/2}$ (D1 line).  The potential curves thus have different slopes, and this changes the effective time for the atom pair to reach the bare atomic asymptote. This implies that a larger fraction of the photons participating in light-assisted collisions under D2 line excitation would be closer to the bare atomic resonance and hence radiation trapped in the gas than for D1 line excitation. Furthermore, while the $0_{g}^{+}$ state contributions to the D2 line and D1 lines are nearly the same, the $1_{u}$ state contribution to the D2 line is significantly larger than that of the D1 line \cite{movre}.  The combination of these effect could explain the significant differences in heating observed experimentally.  Clearly, more theoretical studies on the severe heating with D2 line excitation and the lack thereof with D1 line excitation are warranted.

\vskip 5pt
One of the dominate features of an ultra-cold quantum gas such as a Bose-condensate is its very high degree of de Broglie wave coherence.  From wave propagation theory one can easily understand that radiation-trapped photons will significantly damage the matter-wave coherence. Mathematically, this is expressed as a broadening of the two-photon momentum state transition width which detrimentally impacts the coherent photon generation. To investigate this we further probed the condensate coherence in a series of double-pulse experiments by applying linearly-polarized light pulses along the short axis of the condensate. Figure 4a is a plot of the overall scattering efficiency when a red-detuned pulse overlaps completely with a blue-detuned pulse of variable pump rate. Here both pulses had the same perpendicular polarization, and the blue Rayleigh scattering rates (scattering efficiency) were normalized to the 157 Hz rate (33\% efficiency) of a single red-detuned pulse. Obviously, the heating from blue-detuned light lowers the condensate coherence and depletes the gas. Notice that although this reduces the overall efficiency, it does not completely suppress the scattering. This is because the perpendicularly polarized red-detuned pump provides additional gain to the superradiant process that works in conjunction with the blue-detuned pump to overcome the negative impact of the increased momentum width of the gas (see discussion below) and reach the threshold condition for run-away gain. In Fig. 4b we fixed the red- and blue-detuned pump-pulse durations and scattering rates at 400 $\mu$s and 157 Hz, respectively, and only varied the delay between application of the blue (first) and the red (delayed) pump lasers. As before, the blue-detuned pulse negatively impacted the efficiency, but it could not completely suppress coherent scattering.

\begin{figure}
\centering
\includegraphics[angle=90,width=3.25in]{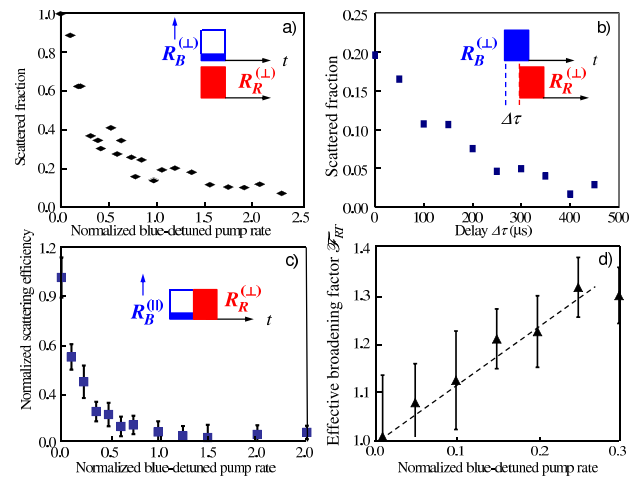} 
\caption{a). Scattering efficiency in a double-pulse experiment as a function of $R$ of an overlapping blue-detuned light. b). Scattering efficiency as a function of the delay time between the blue-detuned and red-detuned pumps. ($R_B^{(||)}=R_R^{(\bot)}$ = 157 Hz). For both plots $\tau$ = 400 $\mu$s, TOF = 15 ms, $\delta/2\pi =\mp 2$ GHz (red/blue).  Efficiency and $R$ are normalized to that of a single red-detuned pulse. c). Scattering efficiency as a function of the parallel polarized blue-detuned rate. d). Effective two-photon resonance broadening factor calculated from c). The dashed line is a guide.}  
\end{figure}
\vskip 5pt
A clearer understanding of the underlying physics can be gained by applying a perpendicularly-polarized red-detuned pump and a parallel-polarized blue-detuned pump at the same time. This allows us to ascertain how the light-assisted collision channels impact the coherently generated optical field that drives the superradiant process. To accomplish this we simultaneously applied a blue-detuned pump with a parallel polarization and a red-detuned pump with a perpendicular polarization.  This is equivalent to replacing $R_{B}^{(\bot)}\rightarrow R_{B}^{(||)}$ and taking $\Delta\tau$ = 0 in Fig. 4b.  Because parallel-polarized light cannot emit photons along the long axis of the condensate, it cannot participate in the superradiant scattering process. Therefore the effect of the parallel-polarized blue-detuned pump was only to induce collisions whose rate we assume to first order is not affected by the orientation of the linear polarization. In contrast to Fig. 4a we found that superradiance did not occur under this orthogonal-polarization, double-pump excitation scheme when the rates $R$ were the same. When the rates are equal, this double-pulse experiment can be viewed as a pseudo-blue pulse because we replaced the blue gain by the red gain while maintaining the same loss rate from photo-induced collisions. 
\vskip 5pt
In Fig. 4c we first applied a 400 $\mu$s parallel-polarized blue-detuned pulse of variable scattering rates and then subsequently applied a 400 $\mu$s perpendicularly-polarized red-detuned pulse with a fixed $R$ = 157 Hz to directly measure the impact of photo-induced collisions on the gas. Basically, the superradiant scattering efficiency of the red-detuned pulse was used as a probe of the detrimental effects induced by the blue-detuned pulse. Figure 4c shows the normalized superradiant scattering efficiency from this double-pulse experiment where the blue rate was varied from 0 Hz to the full 157 Hz rate of the red pulse. As can be seen from the figure, the gain of the superradiant process is largely quenched when the blue rate was only half that of the red. 
Heating from the blue pulse increases the momentum spread of the gas, resulting in broadening of the two-photon resonance line width and reduction of the density, both of which negatively impact the growth of the internally-generated coherent field. Intuitively, the broadening of the two-photon resonance line width can be expressed in terms of an effective two-photon resonance broadening factor $F_{RT}$ such that
$\gamma_{0}\rightarrow F_{RT}\gamma_{0}$ \cite{lu1} 
where $\gamma_{0}$ is the intrinsic two-photon resonance linewidth \cite{stenger}. Figure 4d shows this effective two-photon resonance broadening factor $F_{RT}$ calculated from the measured scattering efficiency using the theoretical coherent wave-propagation gain constant found in Ref. \cite{lu1}. 
Note that beyond a relative pump rate of 0.3, the low signal-to-noise ratio causes the calculated $F_{RT}$ to have a relatively large uncertainty. In addition, depletion effects from heating start to become important, and this introduces yet another efficiency-reduction mechanism.
\vskip 5pt
In conclusion, we presented experimental studies that support the postulation that light-assisted collision and subsequent radiation trapping are the main reasons behind the red/blue detuning asymmetry in matter-wave superradiance. We observed a high atom loss rate and a large momentum spread that cannot be explained by either a simple two-photon Raman scattering process or the intrinsic Rayleigh scattering rate. However, our data are consistent with the effects of heating expected from radiation-trapped photons emitted during a photo-induced collisional process.  Contrary to the experimental results reported in Ref. \cite{muller} we observed severe and comparable heating with all three pump polarizations. With a series of two-color, double-pulse experiments we further demonstrated that the key to understanding the detuning asymmetry does indeed lie in the increased momentum spread of the gas caused by photo-induced collisions. We emphasize that matter-wave superradiance is fundamentally a self-stimulated, wave-mixing process whose efficiency will obviously be impacted by any increase in the momentum width of the gas. Finally, we note that in the case of a single-spin degenerate fermion gas, the Pauli exclusion principle prohibits photo-assisted collisions (molecular formation) and this is why no pump-laser detuning asymmetry was observed experimentally \cite{wang}.  If one creates a two-component spinor fermionic system from a single-spin degenerate fermion gas by RF spin flips, then a scattering asymmetry with respect to the sign of the pump-laser detuning should manifest itself.

Acknowledgments: We thank Dr. Charles W. Clark, Dr. P.S. Julienne, 
Dr. E. Tiesinga,
Dr. P.D. Lett (all at NIST), and Prof. Hui Zhai (TsingHua University) for discussions. Ruquan Wang acknowledges financial support from the National Basic Research Program of China (973 project Grant No. 2006CB921206), the National High-Tech Research Program of China (863 project Grant No. 2006AA06Z104), and the National Science Foundation of China (Grant No. 10704086). 
Technical assistance from Prof. Zhiping Zhong of IOP CAS is also acknowledged.



\begin{thebibliography}{13}



\bibitem{exp} L. Deng et al., Phys. Rev. Lett. {\bf 105}, 220404 (2010).

\bibitem{inouye} S. Inouye et al., Science {\bf 285}, 571 (1999). 

\bibitem{ketterle} W. Ketterle and S. Inouye, C.R. Acad. Sci. Paris t. 2, IV, 339 (2001).

\bibitem{moore} M.G. Moore and P. Meystre, Phys. Rev. Lett. {\bf 83}, 5202 (1999).

\bibitem{li} $\ddot{\rm O}$.E. Müstecaplioglu and L. You, Phys. Rev. A {\bf 62}, 063615 (2000).

\bibitem{piovella} N. Piovella et al., Opt. Commun. {\bf 187}, 165 (2001).

\bibitem{pu} H. Pu, W. Zhang, and P. Meystre, Phys. Rev. Lett. {\bf 91}, 150407 (2003).

\bibitem{bonifacio} R. Bonifacio et al., Opt. Commun. {\bf 233}, 155 (2004).

\bibitem{benedek} C. Benedek and M. G. Benedikt, J. Opt. B: Quantum Semiclass. Opt. {\bf 6}, S111 (2004).

\bibitem{zobay} O. Zobay and G. M. Nikolopoulos, Phys. Rev. A {\bf 73}, 013620 (2006).

\bibitem{yu} Yu.A. Avetisyan and E.D. Trifonov, Laser Physics Letters 1, 373 (2004). 

\bibitem{lu1} L. Deng, M.G. Payne, and E.W. Hagley, Phys. Rev. Lett. {\bf 104}, 050402 (2010). M.G. Payne et al. (unpublished).

\bibitem{muller} While analyzing atom loss rates, heating effects, and the momentum spectra in the preparation of this manuscript, the work of N.S. Kampel et  al., arXiv:1111.6039v1 [cond-mat.quant-gas] (Phys. Rev. Lett. 108, 090401 (2012)) was brought to our attention. 

\bibitem{muller2} We point out that the necessary phase matching conditions in momentum space ($k_P-k_G=K_M$) and in the time domain ($4\omega_R=\Delta_L$, which is the Bragg condition) for coherent growth of the generated field were correctly derived in Ref. \cite{lu1} for a red-detuned pump.  The phase matching condition claimed in \cite{muller} (see N.S. Kampel et  al., arXiv:1111.6039v1 [cond-mat.quant-gas]) does not explicitly lead to either the Bragg condition or momentum conservation.  Furthermore, it requires that the combined momentum and time domain phase be $\pi/2$ for all time and spatial coordinates, which is non-physical.   

\bibitem{lett} Kevin M. Jones et al., 
Rev. Mod. Phys. 78, 483-535 (2006). J. Weiner et al., Rev. Mod. Phys. 71, 1 (1999).

\bibitem{suominen} K.-A. Suominen et al., Phys. Rev. A {\bf 51}, 1446 (1995).

\bibitem{walter} S. Bali, D. Hoffmann, and T. Walter, Europhys. Lett. {\bf 27}, 273 (1994).

\bibitem{burnett} K. Burnett, P.S. Julienne, and K.-A. Suominen, Phys. Rev. Lett. {\bf 77}, 1416 (1996).  

\bibitem{movre} Mladen Movre and Goran Pichler, J. Phys. B: atom. Molec. Phys. 10, 2631 (1977).

\bibitem{stenger} J. Stenger et al., Phys. Rev. Lett. {\bf 82}, 4569 (1999).

\bibitem{wang} P. Wang et al., Phys. Rev. Lett. 106, 210401 (2011).

\end{thebibliography}
\end{document}